 \documentstyle[12pt,a4]{article}
 \author{ Stefano Battiston $^{(1)}$ Eric Bonabeau $^{(2)}$
     and G{\'e}rard Weisbuch $^{(1)}$ \\ %[0.2 cm]
   {\small (1) Laboratoire de Physique Statistique, ENS, 24 rue Lhomond,
            75005 Paris, France}\\
   {\small (2) Icosystem Corp., 545 Concord Av. Cambridge, MA 02138, USA}\\[0.1 cm]
   {\small Correspondence to: stefano.battiston@ens.fr, tel +33144323623, fax +33144323433 }\\[0.1 cm]
 }

 \title{Decision making dynamics in corporate boards}

\begin{document}

 \input epsf
 \maketitle

 \abstract
{ Members of boards of directors of large corporations
 who also serve together on an outside board, form the so called
interlock graph of the board and are assumed to have a strong influence
on each others' opinion. We here study how the size and the topology of
the interlock graph affect the probability that the board approves a
strategy proposed by the Chief Executive Officer.
 We propose a measure of the
impact of the interlock on the decision making, which is found to be a
good predictor of the decision dynamics outcome. We present two models of
decision making dynamics, and we apply them to the data of the boards of
the largest US corporations in 1999. }\newline

  \section{Introduction}
The boards of a set of companies together with their directors form a
bipartite network. The director network is the network obtained taking
the directors as nodes, and a membership in the same board as a link. It
is well known that the director network of the largest companies in the
US and in other countries has a high degree of interlock, meaning the
fact that some directors serve on several boards at the same time so that
many boards are connected by shared directors. Interlock convey
information and power. For example a bank that lends money to an industry
can use interlocked directors in industries of the same domain to get
additional information about the real risk of the loan.
\newline
As a consequence of economic power concentration over the last decades,
"a special social type emerges spontaneously, a cohesive group of
multiple directors tied together by shared background, friendship
networks, and economic interest, who sit on bank boards as representative
of capital in general"[1]\newline% (Mintz and Schwartz, 1985)
 Now, while part of the public opinion has been since long ago
 concerned about the fact that the corporate
elite would represent a sort of "financial oligarchy controlling the
business of the country"[2], stockholders are more concerned about the
effectiveness of boards in overseeing management.
% (Brandeis 1914)
Board's directors should in fact monitor managers's strategies and
decisions to the interest of stockholders. So, after the recent highly
visible cases of bankruptcy in the US, the role of boards in the decision
making process is now largely debated and more sophisticated forms of
corporate control are advocated.
\newline
In this regard there are some works in sociology, investigating whether
boards have adequate knowledge and information to make meaningful
contributions to strategic decision making. Authors try to assess how
multiple boards appointments affect directors' ability to contribute to
strategy [3][4].
%[3]M.A.Carpenter and J.D.Westphal,(2002); [4]E.M.Fich, L.J.White,(2000)
This kind of study is usually done by means of surveys and no modelling
of the dynamics is involved.
\newline
 Some authors have studied the topological properties of the corporate
elite network: Davis and collaborators have shown [5] that the directors
% [5] Davis et al. 2002
network and the boards network of the largest US corporations has Small
World properties [6]. % [6] Strogatz Watts
German firms too, turn out to form a Small World [7].
%[7] German firms...
Vedres [8] has analyzed the social network composed by directors of the
% [8] Vedres
the largest Hungarian companies, banks and government leaders relating
the power of social actors to their properties as nodes of the
network.\newline
 Finally, some authors have studied the diffusion of
governance practices such as the so called 'poison pill' and 'golden
parachute' [9], throughout the board network, with an epidemiological
approach.\newline%[9]Davis and Green 96
In our work we combine the study of the topological properties of the
interlock with the modelling of the dynamics of decision making.
 Directors have to vote in order for the board to take
a decision. It is clear that the social ties linking a director to the
other directors in the board influence the formation of his/her opinion.
Two directors in the board who also serve in another outside board are
likely to take each other's opinion into account more seriously than the
opinion of another director (see below). If there are several directors
that, within a board, share additional ties among them, they form a sort
of "lobby". The question we address here is whether the lobby can
influence significantly the decision making process of the whole board.
The problem concerns of course not only the boards of large corporations,
but many governance structure in social institutions and it is of general
interest in social science modelling. We study the boards of the largest
corporations because it is a relatively well defined framework and there
are data available about the social connections among agents.
\newline
 Now, one can think of two kinds of decisions a board
is faced to: there are decisions regarding a topics specific to a board,
such as the appointment of a candidate member. For such decisions, we
might suppose that different boards don't influence each other. There are
also decisions about topics related to general trends in the economy such
as whether to fire part of the employees, depending on the forecast of
economical recession or whether to adopt some governance practice [9]. In
those cases, decisions previously taken in some boards might influence
other boards. The present work only considers decision of the first kind
when a single board decides on some issue independently of other boards
decisions.

 In general, models of social choice
 assume that agents form their opinion according to the
information available to them about the state of the world and to the
opinions of other agents [10][11][12].  As we said, the interlock comes
in the decision making process because we will assume that two directors
serving at the same time on several boards have stronger influence on
each other. One of the rationale for this assumption is the fact the
recruiting mechanism itself relies on personal familiarity: a candidate
member is proposed and supported by members who already know him/her
because they serve or have served together in another board [4][5]. As a
result, interlocked members are likely to be more influential on each
other's opinion.\newline

There exist a large literature about committees and collective decision
making, but little numerical or analytical modelling. We start from the
standard assumptions of herd behavior [10], which describe individual
decisions as based on the successive surveys of other agent opinions; we
call this first model to be later fully described as a survey model.
 A second model is based on the succession of interventions
of speakers during the board session, each speaker influencing other
directors during his (or her) intervention; this model is called
a broadcast model.

We then investigate, for the two models, the effect of the size and
topology of the graph of interlocked directors of a board, on the final
board decision.
\newline

This paper is organized as follow: we first present some statistics based
on empirical results concerning the US Fortune 1000 companies. We then
describe the survey model, the relevant quantities to be monitored in
simulations and check the simulation results with standard mean field
results in the absence of interlock. The next section is devoted to a
search for a good predictor of the dynamics in the presence of interlock,
and to simulation results obtained with test interlock graphs and with
empirical board interlock graphs. Similar tests are done for the
broadcast models. In the last section we compare and discuss the results.

\section{Interlock graphs}
In the literature, the topological properties of interlocking directorate
are studied for the director network as a whole [5]. We here focus
instead on the interlock inside each single board. We call
\textit{\textbf{interlock graph of a Board}} the graph obtained by
representing directors of a board as nodes and drawing an edge between
two directors if they serve together on an outside board (\textbf{Figure
1}).\newline
 Before investigating how the structure of the interlock
graph affects the decision making process, we want to know how a typical
interlock graph looks like in real boards. We have analyzed data that
have been kindly provided by G.Davis [5], about the boards of the US
Fortune 1000 companies (year 1999). We found that 321 boards out of 821
have a non empty interlock graph. 20 per cent of all boards have a 1-link
interlock graph, another 20 per cent have a more complex interlock graph.
An example of a board with a complex interlock graph, the board of
directors of the Bank of America Corp. is shown in Figure 1. Within the
321 interlock graphs there are chains, cliques ( subgraphs in which each
node is connected to all the others ) and various combinations of these
components. In particular, we looked at the largest clique in the
interlock graph and we found 25 boards with a clique of three nodes, 9
with a clique of 4 nodes. We also looked at the largest connected
component (LCC) in the graph and we found 65 boards with a LCC of 3
nodes, 31 boards with a LCC of 4 nodes, 9 boards with a LCC of 5 nodes, 4
boards with a LCC of 6 nodes and 2 boards with a LCC of 8 nodes.\newline
 We present in \textbf{Figure 2} the histograms of board size (number of
directors in the board), lobby size (number of directors involved in the
interlock graph) and number of links of the interlock graph. Only the 321
boards with non empty interlock graph are considered in the histograms.
The average board size is $12.4\pm3.6$, the distribution is unimodal,
skewed to the right. The smallest and largest board have size 5 and 35
respectively. Lobby size ranges up to 12 nodes. Another interesting
quantity is the ratio between size of the lobby and size of the board
(bottom left) which has a mean of .19. The distribution is obviously non
gaussian with a long tail.\newline From the above analysis we see that
the fraction of boards of the 1000 Fortune companies, that exhibit a
complex interlock graph, is far from being negligible. It is therefore of
great interest to try to model its effect on the decision making
dynamics.

\section{ The survey model}
We want to model the process of decision making on a single board.
 We first consider the most standard model in economics used
to model herd behavior, which we here call the survey model. The model is
basically an iterated voting process.

At each time step, one director randomly chosen
polls opinions of other agents and makes his opinion
accordingly, most often taking the opinion of the majority.

 More precisely, at the board meeting, the CEO proposes a strategy for
the company. The board directors discuss the strategy and at the end take
a decision by voting. We stylize the situation saying that there are only
2 opinions: opinion +1 corresponds to approving CEO's strategy, and -1 to
refusing it. The CEO always sticks to opinion +1. The other directors can
have opinion +1 or -1. Directors discuss between each other and get to
know the opinions of all their colleagues, which they take into account
to formulate a new opinion.
 Other colleagues' opinions define a field:
the field is a weighted sum of colleagues' opinions, where the
weights depend on the number of boards on which two directors sit
together. The new opinion depends stochastically on the intensity of the
field.
\newline In formulas the model reads as follows. The opinion of
director $i$ is a binary variable $s_{i}=\pm1$. The field acting on
director $i$ is:
\begin{equation}\label{}
    h_{i}=\sum_{j=1}^{m}J_{ij}s_{j}
\end{equation}
m being the size of the board, $J_{i,j}$ being the number of boards on
which directors $i$ and $j$ sit together. Obviously, directors take into
account their own opinion, hence $J(i,i)$ in equ. 1 must be non zero.
Setting $J(i,i)$ to 1 is not very realistic, since it implies that a
director with some interlock ties assigns a larger weight to his
colleagues' opinion than to his/her own opinion. We chose to set $J(i,i)$
as the number of boards where director $i$ serve with at least one other
director of the same board.\newline

The probability that director $i$ takes some opinion $\pm1$ at time $t+1$
is given by:
\begin{equation}\label{}
    P\{s_{i}(t+1)=\pm1\}=\frac{ \exp(\pm \beta h_{i}(t))}
                             {\exp(\beta h_{i}(t))+\exp(-\beta h_{i}(t))}
\end{equation}
Parameter $\beta$ in the opinion update acts as the inverse of a
temperature. It measures the degree of independence of a director's
opinion from the field. At T=0 the opinion dynamics becomes
deterministic, at infinite T the dynamics becomes random. The Boltzmann
formalism, often referred to by economists as the logit function, can be
justified by several considerations such as errors in opinion propagation
and random fluctuations of some external conditions. What is meaningful
for us is that a small amount of fluctuation is sufficient to remove the
system from spurious attractors.
\newline
In the next we will refer to dynamics with CEO and without CEO, meaning
respectively, that there is a director with a constant opinion +1, or
not.
\newline Formally the model is analogous to an
Ising magnetic system.

\subsection{Variables characterizing the state of the system}
We here define some macroscopic variables describing the state of the
system. The value of the opinion averaged over directors of the board is
called $M$:
\begin{equation}\label{}
    M \equiv \frac{1}{m}\sum_{j=1}^{m}s_{j}
\end{equation}
$M$, is the analog of the magnetization in the Ising model and is a
function of time. We denote the magnetization at time $0$ and at large
time $T$ as respectively: $M^{0}\equiv M(t=0)$ and $M^{*} \equiv M(t=T)$.
In order to evaluate the impact of the interlock on the decision making
process we consider the probability that the board votes +1 at large time
T:
             \begin{equation}\label{}
                 P_{+}=P\{ M^{*}>0\}
             \end{equation}
If the board is neutral at the beginning i.e. $M^{0}=0$, then in the
absence of interlock and CEO there are equal chances of outcome $M^{*}>0$
or $M^{*}<0$. One way to measure the impact of the interlock is to
consider the probability that the board votes +1, conditional to the
initial average opinion $M^{0}$ in the board being zero:
\begin{equation}\label{}
                 P_{+}^{0}=\{ M^{*}>0 \hspace{.25 cm} | \hspace{.25 cm}
M^{0}=0 \}
\end{equation}
%P_{+}^{0}=\{ M^{*}>0 \hspace{.4 cm} and \hspace{.4 cm} M^{0}=0 \}
The fact that the CEO sticks to opinion +1 can be regarded as a constant
external field: $h_{CEO}=J_{CEO,j}s_{j}=J_{CEO,j}$\newline

\subsection{Dynamics in the case of no interlock}
In the absence of interlock ($J_{ij}=1 \hspace*{.25cm}\forall i=1:N$),
and in the absence of the CEO, the dynamics is equivalent to
ferromagnetism in the mean field approximation.

 The absolute value of $M^{*}$, as a function of beta, shows
a clear phase transition around $\beta = 1$, as predicted by the mean
field theory, even for a small number of directors $N_{d}=10$.

 Another way to visualize the
phase transition is with a bifurcation diagram: \textbf{Figure 3} shows
the probability distribution of values of  $M^{*}$
 obtained in 1000 runs as a
function of beta. When $\beta=0$ one observes a small magnetization due
to the CEO: $M^{*}~\frac{1}{N_{d}}>0$.

 Let us note
that for $\beta>2$, in practice the only possible values of $M^{*}$ are
$\pm1$, so that the probability that $M^{*}$ is positive coincides with
the probability that $M^{*}=1$.\newline
 This means that, at high beta,
the board ends up with deciding at unanimity whether to approve or reject
CEO's proposal (unanimity minus one in case of rejection). For $\beta=4$
the attractor is reached within 25-30 steps. This is a realistic scenario
because typical discussions end up with very large majority in less than
about 3 intervention per director. All the results shown in the following
are obtained with $\beta=4$. In order to compute $M^{*}$, we run the
dynamics for 50 steps and we average the magnetization over time steps
25-50.

\subsection{ Measuring the impact of interlock graphs}
We want to investigate the effect on the dynamics due to the presence of
a group of directors which serve together on one or more outside boards.
The value of $J_{ij}$ is the number of boards on which directors $i$ and
$j$ serve together. Obviously $J_{ij}$ worths at least 1 for all the
directors in the board. The subset of directors of the board for which
$J_{ij}>1$ form a graph, the interlock graph of the board as we called it
in the previous sections. We will refer to it as the 'lobby'. This graph
consists of one or more \textit{Connected Components}(CC). As a result of
the connection structure, a director belonging to a CC feels a stronger
influence from a colleague within the CC than from a colleague outside
the CC. This fact will lead in the following section to a definition of
force of the lobby.\newline \par

We first tested our methodology with the set of
all graphs with 10 directors connected with at most 3 links.
 \textbf{Figure 4} shows
  all the possible graphs obtained with 1,2,3 links,
with a maximum of 2 links per node. They are referred to as interlock
graph 1,...,15. In each box the nodes represent the 10 directors of the
board. The edges represent the ties between directors which serve
together on an outside board. The black node is the CEO. Note that graphs
6 and 8 consist of a fully connected subgraph of 3 directors. In the
latter case the CEO belongs to the graph.\newline For a given initial
condition the outcome of the dynamics can be either 0 or 1, because of
the stochastic nature of the opinion update. Then, in order to estimate
the probability $P_{+}$ that a given board approves the CEO's proposal,
we repeat the dynamics for a large number of runs. We therefore obtain
values of $P_{+}$ for each type of interlock graph, as a function of the
initial magnetization inside and outside the lobby.\newline

Now, in order to compare results for lobbies with different sizes and
topologies, we seek for a scalar quantity that predicts the impact of an
interlock graph. It is clear that the impact of a lobby must depend on
the number of nodes in it and on the number of links.
The number of links alone does not predict the probability of
approving CEO's proposal. The number of nodes alone doesn't do better.
The fact is that with a same number of links one can build a clique or a
chain, so the topological structure must play an important role, too.
Moreover the initial opinions of the directors in the interlock graph
count a great deal, so we need a quantity which can take them into
account.\newline The best predictor we found is the quantity:
\begin{equation}\label{}
    F= \frac{1}{ m^{2} }\sum_{ij\in G}J_{ij}s_{j}(t=0)
\end{equation}
which we call the force.
 This scalar quantity is in fact the intensity of the
field exerted at time $t=0$, by all the directors in the interlock graph
on themselves. The field is normalized with respect to the size m of the
whole board, because we want to estimate the impact of the interlock
graph with respect to the whole board. The same interlock graph will
affect more strongly a small board than a large board.
\newline To have an intuition
of the notion we want to capture, suppose that at time $t=0$ all the
directors of the interlock graph have opinion $+1$, but the board as a
whole has magnetization $M=0$. The stronger the field the interlock
members exert on themselves as compared to the field exerted by the
directors outside the interlock graph, the more chances that the
directors of the interlock graph stick to their initial opinion at $t
\geq 0$. They would then
 act as an
external field driving the board towards positive values of
magnetization (although in principle the directors of the
interlock graph can change opinion at any time; only the
CEO has a fixed opinion).
\newline
 The
force can take several different values according
 to the different initial opinions +1 and -1 in the interlock graph.
 Hence, each graph has a set of possible value of force. For each
value of the force the dynamics has a certain probability to reach the
attractor $M^*=1$ (as we said, for $\beta>2$ whenever $M^*$  is positive,
it is equal to +1).

 \textbf{Figure 5 } displays $P_{+}^{0}$ (the probability of
approving CEO's proposal when the board is neutral at time 0)
as a function of the force of all the interlock graphs.
The fact the $P_{+}^{0}$ is an increasing function of the force was quite
expected, by construction. What was not clear a priori was that lobbies
with different number of links and different topology but with similar
value of the force do have similar value of $P_{+}^{0}$ which means that
the force is a good predictor of the influence of the lobby on the result
of the board decision making. Moreover, it is a linear predictor.

%We also applied the survey model on a set of 100 interlock graph of at
%most 5 nodes. The board size was set to 12 and the CEO was supposed to be
%outside the lobby. This way the configuration with initial null
%magnetization $M_{0}=0$ is always possible, even when all directors in
%the lobby are in favor of CEO's proposal. Results are shown in
%\textbf{Figure 6}.\newline

\subsection{Voting dynamics simulations in real boards}
We ran simulations of the voting dynamics using the interlock graphs that
we found in the real boards. As for the elementary interlock graphs, here
for each board and for each initial condition, we repeat the dynamics a
large number of times in order to estimate the probability $P_{+}$ that
the board will approve CEO's proposal, as a function of the initial
conditions in the lobby.  In \textbf{Figure 6} we show $P_{+}^{0}$
($P_{+}$ conditional to having $M=0$ at time $t=0$) vs the force, for the
real boards of the US 1000 Fortune companies.

To simplify the graph, we considered only points relative to two initial
conditions: apart from the CEO, the directors in the interlock graph
either have all opinions +1 or all -1 at time $0$. Hence each board is
represented by 2 points: one with a positive value of the force and one
with a negative one. \textbf{Figure 6} displays a strong linear
correlation between  $P_{+}^{0}$ and the force.

We left out boards with too large interlock graph i.e. when the number c
of nodes in the interlock graph is larger than half the number m of
directors. In fact in this case, when all the directors in the lobby
start with a same opinion, then, no matter what is the opinion of the
directors outside the lobby, there is no configuration with $M_{0}$ equal
to 0. In this case, we set all directors that are not in the lobby as
against the lobby, but since $M_{0}>0$,  the corresponding data points
are not comparable with the ones of the other boards. Boards with $c>m/2$
(not shown) have $P_{+}^{0}$ values close to 1. \newline

A complementary set of data is obtained by taking the histogram of the
fraction of boards which would agree with the CEO with a given
probability. The set of boards is reduced to boards with an interlock
graph,
 and the initial condition are
$M_{0}=0$ with all lobby members voting initially as the CEO.
\textbf{Figure 9} display these histogram for four sets of simulation
concerning the two models. The top histograms correspond to the survey
model with and without a lobby, for the sake of comparison. One reads the
histograms in the following way: with the survey model for example (top
right frame), 25 per cent of the boards have 75 per cent chances to
approve CEO's proposal, if the directors in the lobby are initially in
favor of it. Moreover one can say that 40 per cent of the boards have at
least 75 per cent chances of approving the CEO.

\section{The broadcast model}
We consider now a different model for the voting dynamics, based on the
idea that at each time step one director takes his turn to speak while
the other directors listen to him/her and are influenced by his/her
opinion. We will refer in the following to this model as the broadcast
model.\newline

At the board meeting, the CEO proposes a strategy for the company. Again
this is stylized saying that there are only two opinions: opinion $+1$
corresponds to approving CEO's strategy, and $-1$ to refusing it. The CEO
always sticks to opinion $+1$. The other directors can have opinion $+1$
or -$1$.\newline

\begin{itemize}
  \item One director $j$ at a time is chosen to speak.
His own opinion is evaluated, as usual, based on the field he experiences
according to the logit equation (equ. 2).
  \item Only the individual field evaluation is changed.
 When director $j$ speaks, all directors $i$ update their
individual field according to:
\begin{equation}\label{}
    h_{i}^{new}=(1-\gamma)h_{i}+\gamma J_{ij}s_{j}
                \hspace*{.5cm}\forall i
\end{equation}
$\gamma$ is a parameter which determines the memory length of the agent.
At the beginning, the field of the agent $i$ is initialized as equal to
$J_{ii}s_{i}$.
\end{itemize}
As a result, the field experienced by an agent, only takes into account
the discounted opinion of the other agents, at the time when they spoke
(which might be different from their actual opinion now).  This scheme is
closer to a class of models based on the P\`{o}lya urn, also used by
economists[13].% [13] Blackwell,1964.
We might then expect some sensitivity to the ordering of agents'
interventions during the board meeting.

 In fact, the
broadcast model requires to choose at each time step who is going to
speak. As modelers we are tempted to use a random order,
but in real boards the order is probably far from being random:
 more convinced directors will likely try to speak first, and
moreover the CEO or the chairman plays a role in deciding the order of
the speakers. In order to understand the impact of the way in which
directors are chosen to speak, two extreme strategies are investigated
here:

\begin{enumerate}
 \item Strategy 1. The speaker is chosen randomly.
 \item Strategy 2. For $t\leq c$ ($c$ being the size of the interlock graph),
 the speaker belongs to the interlock graph. For $t> c$ the
speaker is chosen randomly.
\end{enumerate}

We have run simulations of the broadcast dynamics on the elementary
interlock graphs and on the real boards of the US 1000 Fortune companies.
Similar results were observed for different $\gamma$ values ($\gamma =
0.1$, $0.3 $). We performed the same analysis as for the survey model: we
estimated the probability $P_{+}^{0}$ that the board will approve CEO's
proposal, conditional to having $M=0$ at time $t=0$, as a function of the
initial conditions in the lobby. As before only the two extreme cases for
which the directors in the lobby are all in favor of the CEO, or all
against him/her are taken into account.
 $P_{+}^{0}$ versus the force is shown in \textbf{Figure 7,8}
for the two strategies of choosing the speakers. The histograms of the
fraction of boards which would agree with the CEO with a given
probability are shown in \textbf{Figure 9}.

\section{Discussion}
We have investigated the impact of different structures of corporate
directors interlock on the outcome of the decision making process of
boards of directors. We have considered two models of decision making
process, and we have studied the probability of board approval of the
CEO's strategy as a function of the topology and the size of the
interlock structure. We have applied the models on a set of test
interlock graphs or lobbies, in order to find a good predictor of the
interlock impact, and then we have applied the models to the boards of
the largest US corporation.\newline

\textbf{Figure 10} summarizes our results: the existence of a lobby does
influence the vote as compared to the absence of a lobby. The probability
$P_{+}^{0}$ that the board approves CEO's proposal when the board is
initially neutral, is plotted against the force of the lobby for the
different models ( for the purpose of comparison, values of the force are
grouped in bins of width 0.05 and the corresponding values of $P_{+}^{0}$
for different boards inside the bin are averaged together). What
surprised us is that this influence is of comparable magnitude for the
survey and random broadcast models, at least for small values of
$\gamma$, the time discount factor. In the broadcast model when directors
of the lobby speak first, the influence of the lobby is enhanced, and
even more so in the neighborhood of zero force. This means that a
strategic sequence of interventions may enhance the power of the lobby on
the decision making process. The discontinuity at $F=0$ increases with
$\gamma$.\newline

We have focused our attention on the case in which initially the whole
board is neutral about the decision, that is $M_{0}=0$, while the
directors in the lobby have the same opinion, either all +1 or all -1. In
this case the probability of approval is related to the power of the
lobby, where "power" is used in accordance to Weber's definition:
 "power of an actor in a social
network is the probability that this actor will carry on his/her will
despite resistance of the other actors"[14].

The interest of this investigation for the social sciences consists in
offering a framework in which it is possible to make quantitative
predictions about the power of a lobby within a board: given the topology
of the social ties, we can compute a quantity, the force, which is a good
predictor of the power of the lobby. In principle the board should take
decisions on the interest of all investors, based on the available
information. From our results, a well connected lobby of a minority of
directors can drive the decision of the board, and the chances that the
board will finally agree with the lobby can be predicted measuring the
force of the lobby.

Having a powerful lobby inside the board
simply means that the opinion of some
directors has counted more than the opinion
of others,  which is not necessarily bad if, for example, the directors
in the lobby were the most competent about the matters in discussion.

But suppose now the lobby rather represents
 the interest of some minority. This
minority could consist of officers of the company itself, reluctant to a
change of management or officers of another company that owns a minority
of stocks and want to attack the company. This could be seen as a
dangerous situation for the company and the majority of investors.
 In this perspective, norms could be introduced
to limit the force of the lobby e.g.
when a new director is proposed for an appointment in the
board.\newline

Of course, the prediction of the outcome of the decision making process
assumes some simple hypotheses about the influence of board directors on
each other's opinion. The main hypothesis of our models is that the
influence $J_{ij}$ of a director $i$ on another director $j$ is a linear
function of the number $n_{ij}$ of boards on which the two directors
serve together. Some different
functional relationships between $J_{ij}$ and $n_{ij}$ could be assumed,
provided that the influence is a monotonic increasing function. There is
no a priori justification for our linear choice, other than the fact that
it is a simple approximation to start with.\newline

For our models, we do not have an estimate of the real value of $\beta$.
We made simulations for different values of $\beta$, then we focussed on
the case $\beta=4$, in order to avoid meta-stable states. But for any
value $\beta>>1$ the dynamics converges very rapidly to unanimity, and
for $\beta=4$ in less than 30 steps, i.e. after about 3 interventions on
average of each director. This is a quite realistic scenario: in fact
this is the typical number of interventions for a board discussion.
Moreover, a typical discussion ends up with a consensus or a large
majority.\newline

The two models that we have investigated differ in the
mechanism of opinion update. The opinion update mechanism we adopted for
the survey model is analogous to what is known as "herd behavior" in the
literature of opinion dynamics, but it is also analogous to what is
well-known in statistical physics as magnetic system dynamics at finite
temperature. In the broadcast model we propose a more realistic mechanism
of opinion update, which takes into account the fact that in a real
discussion one is not informed of everybody else's opinion at each step
in time. Instead, participants speak once at a time, so that each agent
only knows the opinion that another agent had at a certain time, which
may differ from the opinion he has at the current time.\newline

 The present study focuses on boards of directors,
 because of the availability of empirical data.
Our conclusions can be also applied to the decision
 dynamics of any political
committee or academic board.\newline

\par One possible extension of this investigation is the study,
now in progress, of the dynamics of the decision making process of boards
when the decision taken at one board influences the decision process of
other boards. In fact, in the case of discussions about adoption of
governance practices [9] or decisions that require prior forecasting of
economic trends, directors of a board are likely to take into account
decisions made in interlocked boards.

\section{Acknowledgements}
We would like to thank Gerald Davis of the University of Michigan,
Business School, for having kindly provided the data of the US Fortune
1000. We also thank Jacques Lesourne for fruitful discussion.\newline
Eurobios company supported two of us, SB and EB, in the early stage of
this study. We acknowledge the support of the FET-IST department of the
European Community, Grant IST-2001-33555 COSIN.

\section{References}
\begin{enumerate}
%1 Mintz
\item B.Mintz, and M.Schwartz, The Power Structure of American Business,(1985),
University of Chicago Press.
%2 Brandeis
\item L.D.Brandeis, Other People's Money: And How the Bankers Use It.
  New York (1914), Frederick A. Stokes.
%3 Carpenter
\item M.A.Carpenter and J.D.Westphal, The strategic context of social
network ties: examining the impact of director appointment on board
involvement in strategic decision making, forthcoming in Academy of
Management Journal.
%4 Fich
\item E.M.Fich, L.J.White,(2000),Why do CEO's reciprocally sit on each other's
            boards?, forthcoming
%5 Davis et al
\item G.F.Davis,M.Yoo,W.E.Baker,The small world of the corporate elite, 2002, forthcoming.
%[6] Strogatz
\item J.D.Watts and S.Strogatz, Collective Dynamics of "Small World" Networks,
 Nature 393 (1998) 440-442.
%7
\item B.Kogut and G.Walker, The Small World of Firm Ownership in Germany:
Social Capital and Structural Holes in Large Firm Acquisitions -
1993-1997, (1999)
% [8] Vedres
\item B.Vedres, The constellation of economic power: the position of political
actors, banks and large corporations in the network of directorate
interlock in Hungary, INSNA (1997) 23(1):44-59 2000
%[9]Davis and Green 96
\item J.Davis and H.R.Greve, Corporate elite networks and governance changes
in the 1980s, Am. J. of Sociology 103 (1996) 1-37.
% [10] Orl\'{e}an A. (1995)
\item A.Orl\'{e}an, Bayesian interactions and collective dynamics of
opinions: herd behavior and mimetic contagion, J. of Economic Behavior
and Organization 28 (1995) pp.257-274.
%[11] Stauffer
\item D.Stauffer, Monte Carlo simulations of Sznajd models,
J. of Artificial Societies and Social Simulation vol.5 n.1
(2002),\newline http://www.soc.surrey.ac.uk/JASSS/5/1/4.html
%[12] Galam
\item S.Galam,J-D.Zucker, From individual choice to group decision
making, Physica A 287 (2000) 644-659.
%[13] Polia
\item D. Blackwell, D.Kendall, The martin boundary of P\'{o}lya urn
schemes...   J. Appl. Probability 1 (1964) 284.
%[14] Weber
\item M. Weber, Theory of Social and Economic Organization, 1920
% \item K.Sznajd-Weron and J.Sznajd, Int.J.Mod.Phys. C11 (2000), 1157.

\end{enumerate}

\section{Figures}

\begin{figure}[tbh]
    \centerline{ \epsfxsize=100mm\epsfbox{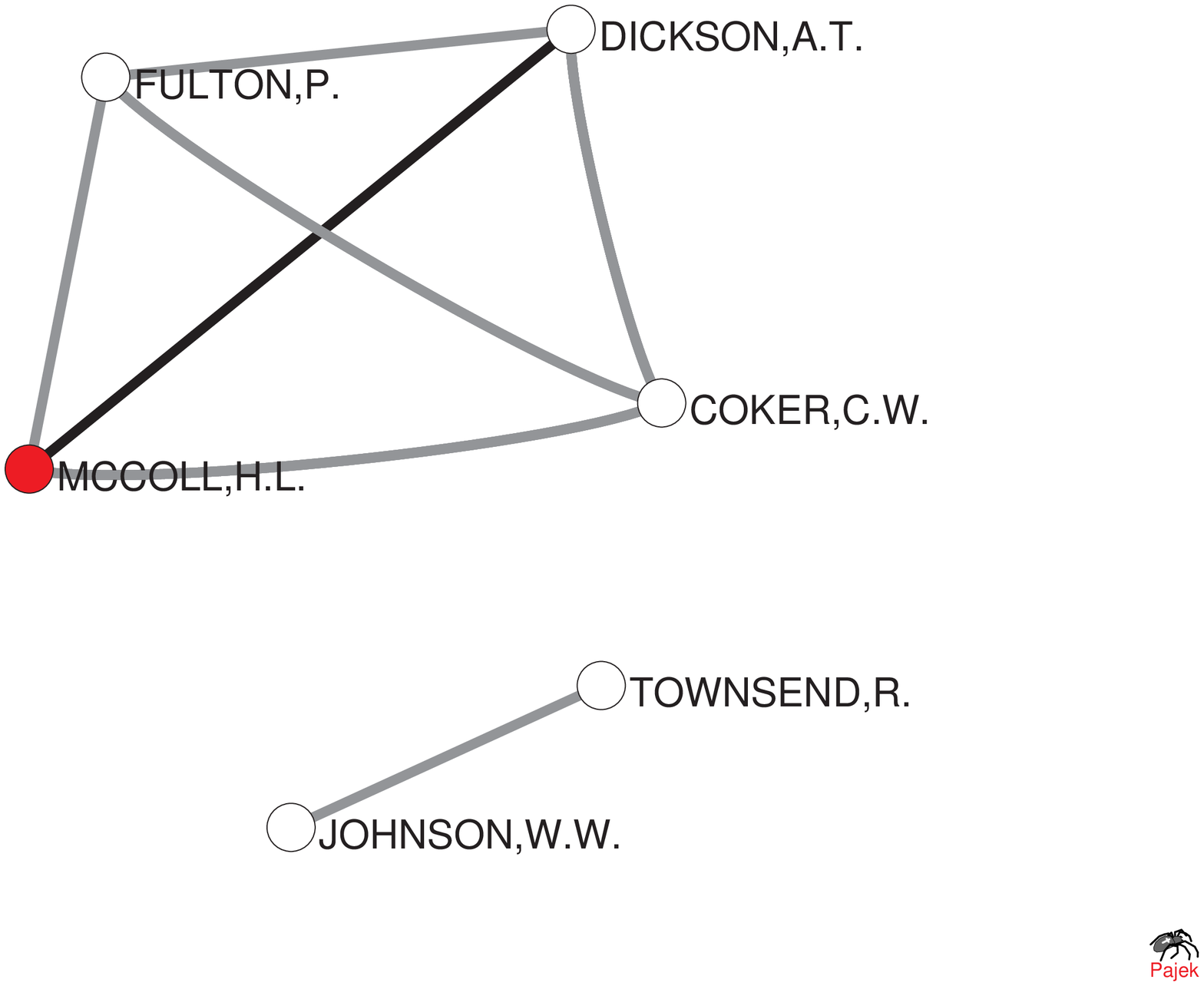}}
    \centerline{\parbox{320pt}{\caption{\label{ba}:
        Example of an interlock graph: The
        board of directors of the Bank of America Corporation.
        White nodes represent directors that are not in the management,
        black nodes represent directors that are also executive of the
        company.
        Two directors are connected by a gray edge when they serve on one
same
        outside board. The edge is black when they serve together on
        more than one outside board.
    }}}
\end{figure}
\begin{figure}[tbh]
    \centerline{ \epsfxsize=100mm\epsfbox{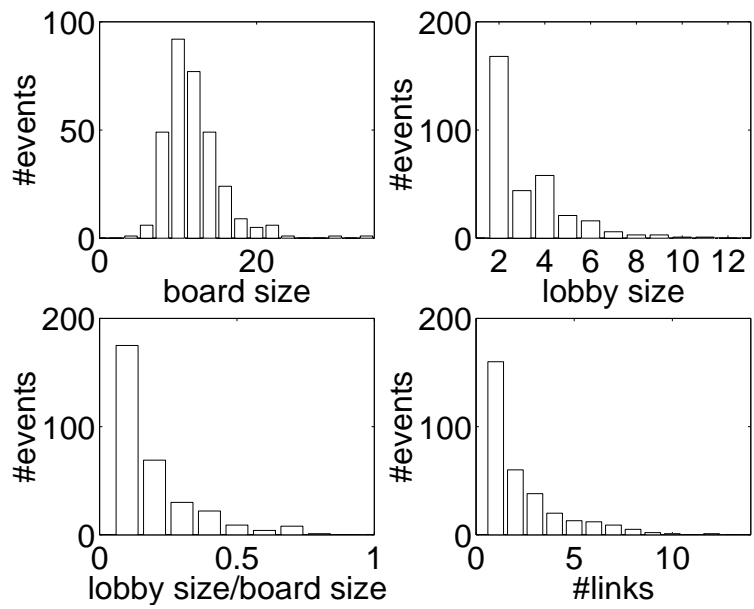}}
    %\centerline{ \epsfxsize==100mm\epsfbox{chase.eps}}
    \centerline{\parbox{320pt}{\caption{\label{ba}:
    Histograms of board and interlock characteristics.
            Top left: board size (number of directors in the board).
            Top right: lobby size(number of directors involved in an interlock
                        tie with some other directors of the same board).
            Bottom left: ratio between lobby size and board size.
            Bottom right: number of links in the interlock graph.
    }}}
\end{figure}
\begin{figure}[tbh]
    \centerline{ \epsfxsize=80mm\epsfbox{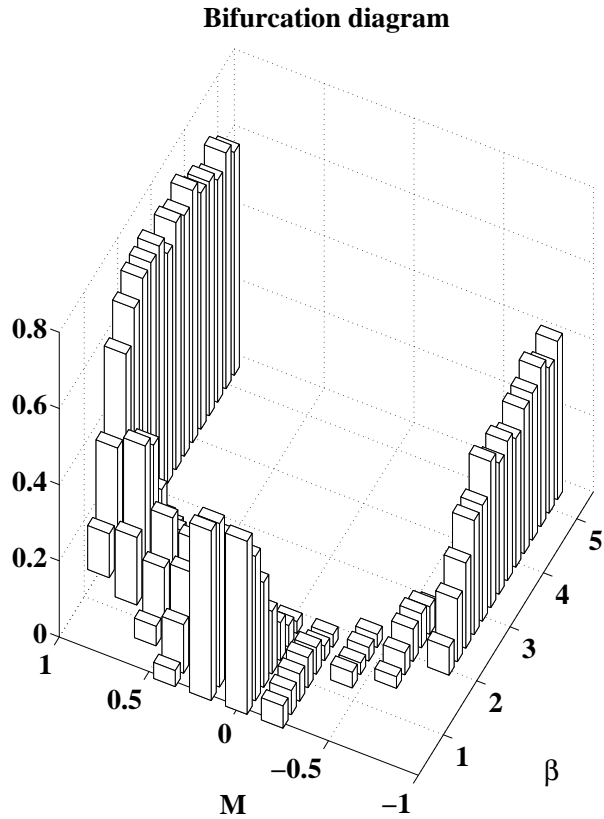}}
    \centerline{\parbox{320pt}{\caption{\label{ba}
      Bifurcation diagram of the magnetization.
      Frequency distribution of the
 final value of the magnetization $M^{*}=M(t=T)$,
      obtained in 1000 runs, as a function of $\beta$,
the temperature parameter.
      The board has 10 directors.
      The asymmetry of the diagram
is due to the presence of the CEO who always vote +1.
      Note that for $\beta>2.5$ the probability that $M^{*}=1$
      coincides with the probability that $M^{*}>0$.
     }}}
\end{figure}

\begin{figure}[tbh]
    \centerline{ \epsfxsize=130mm\epsfbox{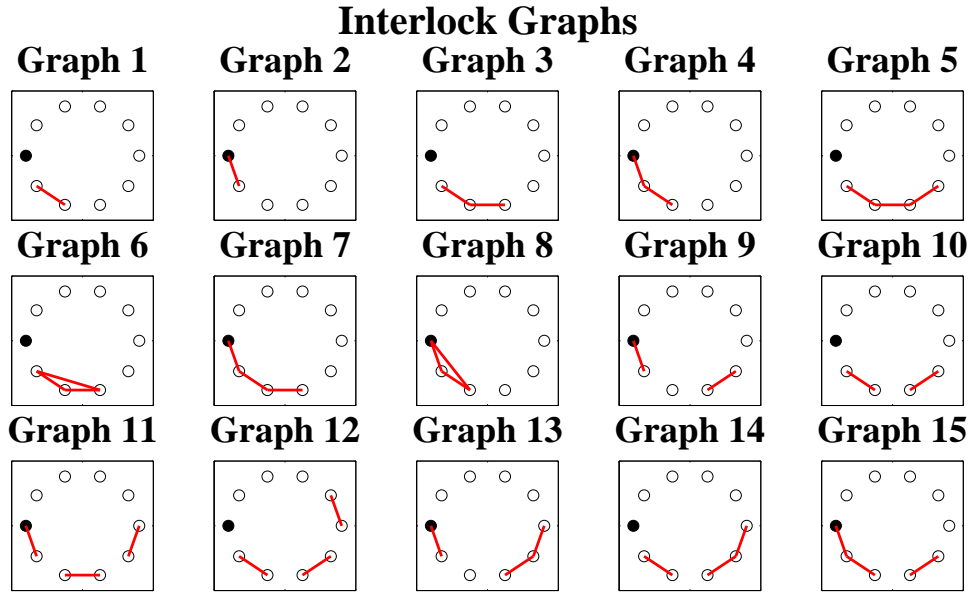}}
    \centerline{\parbox{320pt}{\caption{\label{ba}: The simplest
interlock graphs for boards with 10 directors.
    There are 15 different graphs that can be drawn
    with a maximum of 3 links and with up to 2 links per node.
    The black node is the CEO.
    }}}
\end{figure}

\begin{figure}[tbh]
    \centerline{ \epsfxsize=100mm\epsfbox{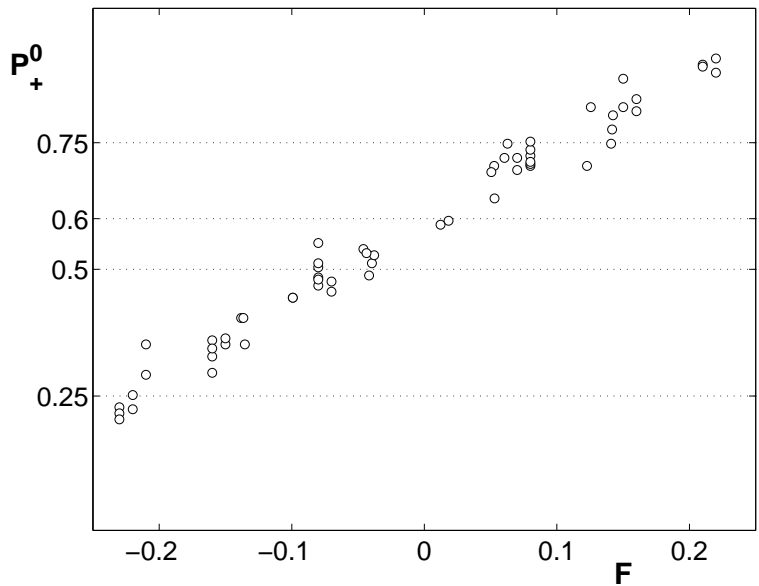}}
    \centerline{\parbox{320pt}{\caption{\label{ba}: Survey model
simulation
        of the 15 elementary interlock graphs.
        Ordinate: probability $P_{+}^{0}$ that the board approves CEO's
        proposal, conditional to the board being initially neutral (
        $M^{0}=0$). Abscissa:  force of the interlock graph.
        Data points are the average of 500 runs. Each data point
        corresponds to a given initial magnetization value of a single
        interlock graph.
   }}}
\end{figure}

\begin{figure}[tbh]
    \centerline{ \epsfxsize=100mm\epsfbox{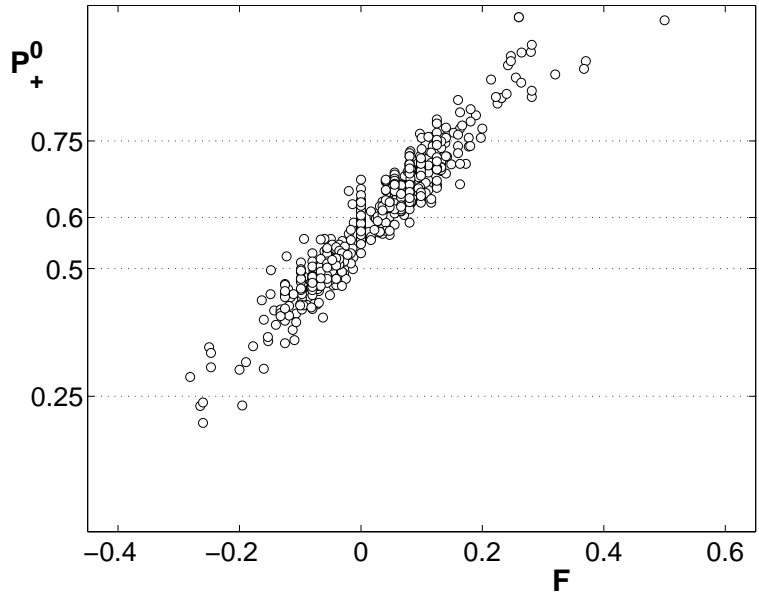}}
    \centerline{\parbox{320pt}{\caption{\label{ba}
     Survey model simulation on the boards of the Fortune 1000 companies.
     Ordinate and abscissa as in figure 5, each point is an average
over 500 runs.
     Only 2 initial magnetization values of the lobby are considered
     ( all +1, all -1).
 }}}
\end{figure}

\begin{figure}[h]
    \centerline{ \epsfxsize=100mm\epsfbox{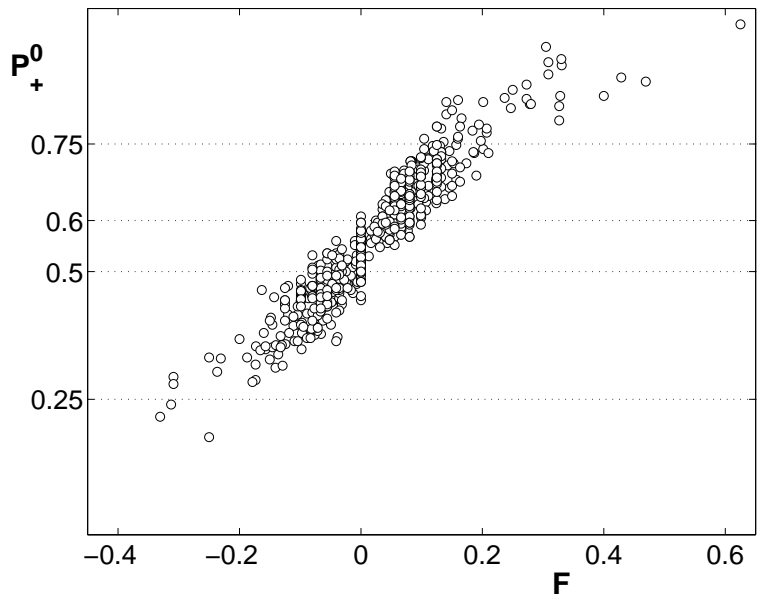}}
    \centerline{\parbox{320pt}{\caption{\label{ba}
    Broadcast model simulation on the boards of the Fortune 1000
companies.
    The order of the speakers is random.
    Ordinate and abscissa as in figure 5, each point is an average
over 500 runs. Only 2 initial magnetization values of the lobby are
considered ( all +1, all -1).
   }}}
\end{figure}

\begin{figure}[h]
    \centerline{ \epsfxsize=100mm\epsfbox{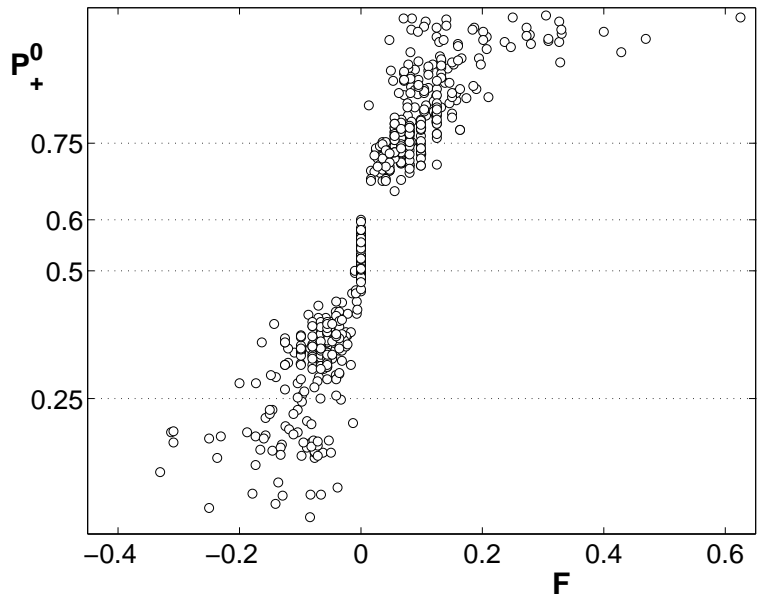}}
    \centerline{\parbox{320pt}{\caption{\label{ba}
    Broadcast model simulation on the boards of the Fortune 1000
companies. The order of the speakers is as follows: directors of the
lobby speak first, then the speaker is chosen randomly.
    Ordinate and abscissa as in figure 5, each point is an average over 500 runs.
     Only 2 initial magnetization values of the lobby are considered
     ( all +1, all -1).
   }}}
\end{figure}

\begin{figure}[tbh]
    \centerline{ \epsfxsize=110mm\epsfbox{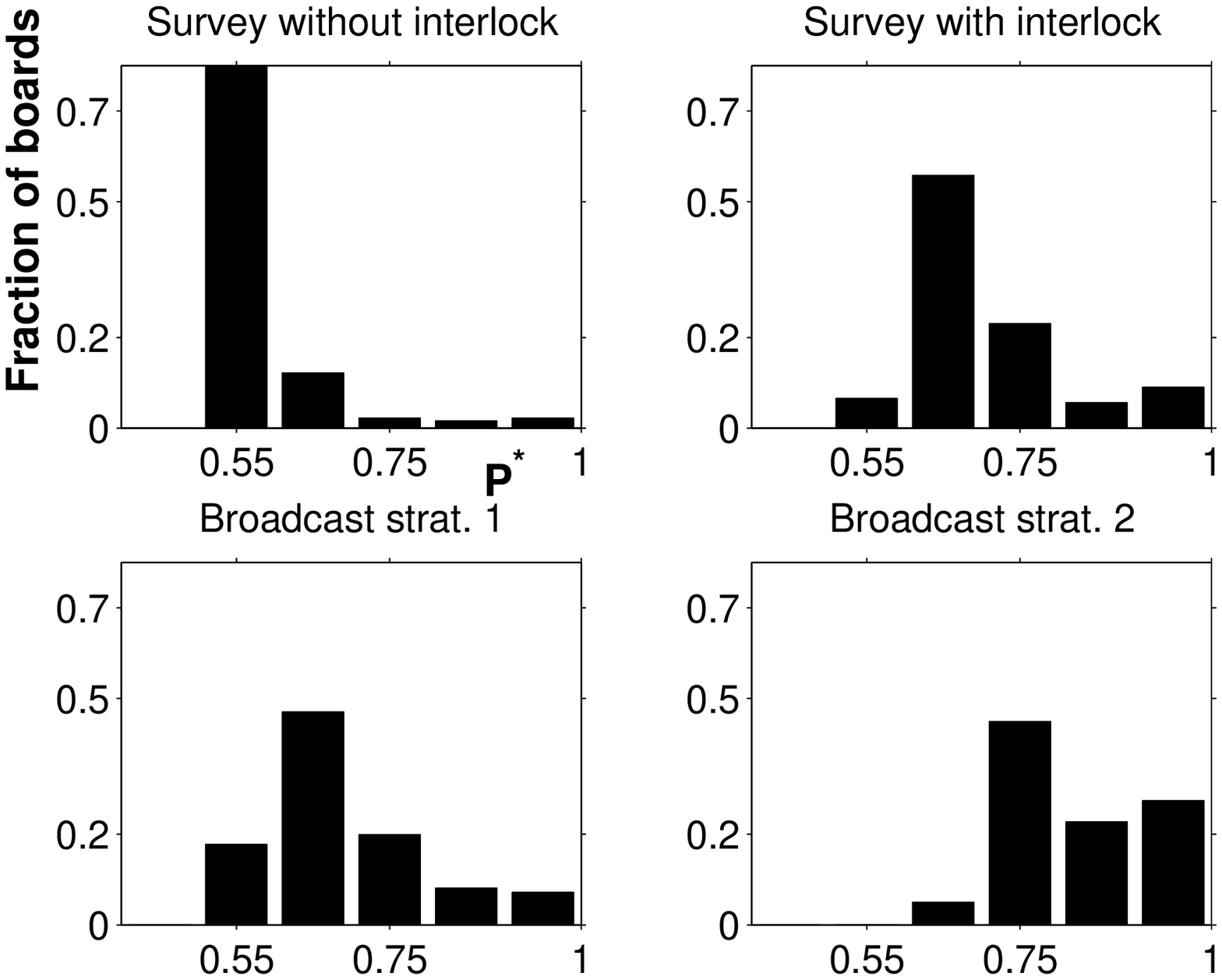}}
    \centerline{\parbox{320pt}{\caption{\label{ba} : Histograms of
the fractions of boards which would approve the CEO with a probability
given in abscissa.\newline
    Top left: values for the survey model without interlock.\newline
    Top right: survey model.\newline
    Bottom left: broadcast model, strategy 1 (random order of
speakers).\newline
    Bottom right: broadcast model, strategy 2 (directors in the lobby
speak first).\newline
   }}}
\end{figure}

\begin{figure}[h]
    \centerline{ \epsfxsize=120mm\epsfbox{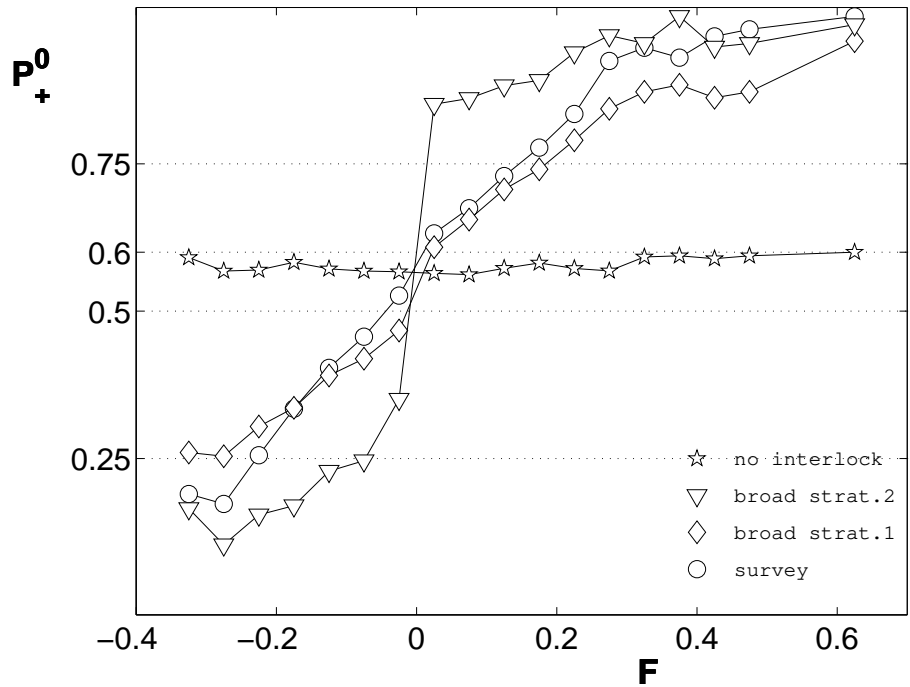}}
    \centerline{\parbox{320pt}{\caption{\label{ba} :
        Comparison of the results of the different models.
        Ordinate and abscissa as in figure 5.
        Data points  correspond to
 the 321 boards with non-empty interlock graph.
        Pentagons: survey model with no interlock for control purposes.
        Circles: survey model.
          Diamonds: broadcast model with $\gamma = 0.1$, strategy 1 (random order of
speakers).
        Triangles: broadcast model with $\gamma = 0.1$, strategy 2 (directors in the lobby
speak first).
   }}}
\end{figure}

\end{document}